# Copper-based charge transfer multiferroics with a $d^9$ configuration

Hu Zhang*, RuiFeng Zhang, Lulu Zhao, Chendong Jin, Ruqian Lian, Peng-Lai Gong, RuiNing Wang, JiangLong Wang, and Xing-Qiang Shi

Key Laboratory of Optic-Electronic Information and Materials of Hebei Province, Research Center for Computational Physics, College of Physics Science and Technology, Hebei University, Baoding 071002, P. R. China

* E-mails: zhanghu@hbu.edu.cn

Multiferroics are materials with a coexistence of magnetic and ferroelectric order allowing the manipulation of magnetism by applications of an electric field through magnetoelectric coupling effects. Here we propose an idea to design a class of multiferroics with a $d^9$ configuration using the magnetic order in copper-oxygen layers appearing in copper oxide high-temperature superconductors by inducing ferroelectricity. Copper-based charge transfer multiferroics $SnCuO_2$ and $PbCuO_2$ having the inversion symmetry breaking *P*4*mm* polar space group are predicted to be such materials. The active inner *s* electrons in Sn and Pb hybridize with O 2*p* states leading the buckling in copper-oxygen layers and thus induces ferroelectricity, which is known as the lone pair mechanism. As a result of the $d^9$ configuration, $SnCuO_2$ and $PbCuO_2$ are charge transfer insulators with the antiferromagnetic ground state of the moment on Cu retaining some strongly correlated physical properties of parent compounds of copper oxide high-temperature superconductors. Our work reveals the possibility of designing multiferroics based on copper oxide high-temperature superconductors.

## I. INTRODUCTION

Ferroelectrics and high-temperature superconductors are two types of attractive materials in condensed matter physics [1,2]. One of the most interesting characters of a ferroelectric material without the inversion symmetry is the switchability of its electronic polarization by an applied electric field. It is also possible that ferroelectrics have magnetic order, which are known as multiferroics [1]. The magnetoelectric coupling effect in multiferroics allows the controlling of magnetism by applications of an electric field. On the other hand, for copper-based high-temperature superconductors, one way to understand the physics is from the doping of a Mott insulator with the antiferromagnetic (AFM) ground state of the moment on Cu [2]. Motived by this, we propose an idea to design a class of multiferroics with a $d^9$ configuration based on copper oxide superconductors using magnetic order in copper-oxygen layers. If this is achieved, multiferroics obtained in this way may retain some strongly correlated physical properties of the parent compound of copper oxide superconductors. The key issue is to induce ferroelectric order in copper-oxygen layers by some physical mechanisms. Rich physics should emerge in such undoped and doped copper oxide multiferroics from the view of fundamental scientific interest. These multifunctional materials with magnetoelectric coupling effects might also have wide potential technological applications.

The parent compound of copper-based high-temperature superconductor found by Bednorz and Müller in 1986 is $La_2CuO_4$ with a layered perovskite structure containing the $CuO_6$ octahedra [3,4]. The "infinite layer" material $CaCuO_2$ is also the parent compound of high-temperature superconductors [5]. The crystal structure of $CaCuO_2$ has a space group *P*4/*mmm* (No. 123) with planar $CuO_2$ planes separated by Ca atoms, as shown in Fig. 1(a). Compared to $La_2CuO_4$, $CaCuO_2$ has a simpler atomic structure. Like $Ca^{2+}$ ion, group IV elements Sn and Pb can also form divalent cations. In addition, we also know that Pb plays an important role in the formation of the ferroelectricity in some ferroelectrics including $PbTiO_3$ [1]. Thus, we may use Sn and Pb to obtain copper oxide multiferroics based on the crystal structure of $CaCuO_2$.

In this work, we design multiferroics starting from the crystal structure of $CaCuO_2$. The obtained multiferroics $SnCuO_2$ and $PbCuO_2$ have a polar space group *P*4*mm* (No. 99) with buckled copper-oxygen layers separated by Sn and Pb atoms respectively. Our theoretical results reveal that the active *s* electrons in Sn and Pb hybridize with O $2p$ states and thus leads a distortion in the copper-oxygen layers breaking the inversion symmetry, which results in ferroelectric order in SnCuO$_2$ and PbCuO$_2$. This is the well-known lone pair mechanisms [1,6,7]. The crystal structure of the ferroelectric state SnCuO$_2$ is shown in Fig. 1(b). The ferroelectric structural transition is from the nonpolar high-symmetry paraelectric state with the *P*4/*mmm* structure into the polar ferroelectric state with the *P*4*mm* structure. In other words, SnCuO$_2$ and PbCuO$_2$ in the CaCuO$_2$ structure are paraelectric states. More importantly, the Cu $3d_{x^2-y^2}$ bands in both SnCuO$_2$ and PbCuO$_2$ cross the Fermi energy, like CaCuO$_2$ [8]. The Heyd-Scuseria-Ernzerhof (HSE) hybrid density functional results give AFM ground state of the moment on Cu sites in two ferroelectrics. The buckling CuO$_2$ plane is a charge transfer insulator. Due to the coexistence of ferroelectric and magnetic order, SnCuO$_2$ and PbCuO$_2$ are multiferroics. Therefore, charge transfer multiferroics are successfully designed from copper oxide high-temperature superconductors. As expected, the designed multiferroics in this work retain some strongly correlated physical properties of high-temperature superconductors.

## II. METHODOLOGY

First-principles calculations were performed based on density functional theory (DFT) [9] with the local density approximation (LDA) [10] in the Vienna Ab Initio Simulation Package (VASP) [11-13]. Strongly correlated electronic structures were calculated with the HSE hybrid functional [14]. We used a 12×12×12 Monkhorst-Pack grid [15] in LDA calculations and a 6×6×6 Monkhorst-Pack grid in HSE calculations. We used an energy cutoff of 500 eV. All crystal structures were relaxed until the Hellmann-Feynman forces are less than 1 meV/Å. The Berry phase method [16] was used to calculate the electric polarization. Tight-binding models were constructed with the maximally localized Wannier functions using Wannier90 package [17,18]. Phonopy was sued to calculate the phonon spectra [19].

## III. RESULTS AND DISCUSSION

### A. Ferroelectric properties of SnCuO$_2$ and PbCuO$_2$

Firstly, we consider the crystal structure of CaCuO$_2$, that is the parent compound of high-temperature superconductors, has the *P*4/*mmm* space group with Ca, Cu, O1, and O2 atoms site at the 1*d* (1/2, 1/2, 1/2), 1*a* (0, 0, 0), and 2*f* (0, 1/2, 0) (1/2, 0, 0) Wyckoff positions respectively. The theoretically lattice parameters are $a = b = 3.77$ Å, $c = 3.10$ Å from our LDA structural relaxation, which are consistent with experimental values of $a = 3.86$ Å, $c = 3.20$ Å [5,8]. As shown in Fig. 1(a), $CaCuO_2$ has a centrosymmetric structure with flat copper-oxygen layers and thus surely there is no ferroelectric order in it. To design multiferroics containing copper-oxygen layers, we should destroy the flatness of copper-oxygen layers leading a broken of the inversion symmetry. From knowledge of ferroelectrics, we know that active inner $s$ electrons in $Pb^{2+}$ and $Sn^{2+}$ ions play an important role in the formation of the ferroelectricity in some ferroelectrics (e.g., $PbTiO_3$). Hence, we try to use Pb and Sn to break the inversion symmetry in compounds containing copper-oxygen layers. In the designed "infinite layer" materials, copper-oxygen layers are separated by Pb and Sn atoms. As we will see below, copper-oxygen layers in such materials lost their flatness as required.

We use the crystal structures of $CaCuO_2$ as the initial input data to obtain the theoretical values for $SnCuO_2$. The relaxed lattice parameters are $a = 3.88$ Å, $c = 3.21$ Å. Then we calculate frequencies of phonon at the Brillouin zone center to check its dynamical stability. There are two phonon modes with imaginary frequencies. One is the infrared active $E_u$ modes with imaginary frequencies of 87 cm$^{-1}$, which is related to displacement eigenvectors in the *x-y* plane. The other one is the infrared active $A_{2u}$ mode with imaginary frequencies of 186 cm$^{-1}$. This phonon mode is polarized along the *z* direction with the calculated displacement eigenvector of [δ(Sn) = +0.524, δ(Cu) = −0.123, δ(O) = −0.595].

According to knowledge of the soft-mode theory of ferroelectrics [1], the unstable $A_{2u}$ polar phonon in $SnCuO_2$ with the *P*4/*mmm* structure is usually related to a ferroelectric transition. Hence, we freeze this phonon to try to transform the nonpolar *P*4/*mmm* state into a dynamically stable polar ferroelectric state. Displacements are added to atoms in the nonpolar *P*4/*mmm* $SnCuO_2$ according to the calculated displacement eigenvectors. Then, the new structure is fully relaxed. The LDA calculations give $a = 3.78$ Å, $c = 3.98$ Å. Sn, Cu, O1, and O2 atoms site at the $1b$ (1/2, 1/2, 0.7016), $1a$ (0, 0, 0.0771), and $2c$ (0, 1/2, 0.9926) (1/2, 0, 0.9926) Wyckoff positions respectively. Structural analyses indicate that this state is a polar phase with the polar space group $P4mm$. Compared to the nonpolar state, the lattice constant $c$ of the polar state increases significantly. Fig. 1(b) shows the crystal structures of the polar state $SnCuO_2$. The relative displacements among Sn, Cu, and O along the $c$ direction are evident. Now Sn is not located at center of the cell. The bond lengths of Sn-O are 2.519 Å and 2.217 Å for nonpolar and polar $SnCuO_2$ respectively.

In Fig. 1(c) we show the top and side view of the copper-oxygen layers in polar $SnCuO_2$. The bond length of Cu-O is 1.918 Å. The Cu-O-Cu bonding angle is about 159.76° due to the buckling of the $CuO_2$ plane. A buckling parameter $d$ can be defined as the distance between Cu and O in the buckled plane along the $c$ direction. For $SnCuO_2$, $d$ is 0.32 Å. This is a very large value. Therefore, Sn induces a significant buckling and thus breaks the inversion symmetry. As a result, there is electronic polarization in the copper-oxygen layers. The physical origin of this structure distortion will be discussed below. The phonon spectra of polar $SnCuO_2$ are shown in Fig. 2(a). There are no modes with imaginary frequencies in the entire Brillouin zone, which indicates the dynamical structural stability of the polar $SnCuO_2$. In Fig. 2(b) we plot the characteristic ferroelectric double-well energy of $SnCuO_2$ as a function of polar distortion obtained by linear interpolation between the nonpolar $P4/mmm$ and polar $P4mm$ structures. The calculated energy barrier is about 0.4 eV, which is in the range favorable for ferroelectric switching [1]. We conclude that the polar phase $SnCuO_2$ with the $P4mm$ space group is a ferroelectric material with the $P4/mmm$ state as the nonpolar high-symmetry paraelectric state.

Next, we consider Pb that also has lone pair electrons $6s$ [7]. The fully relaxed lattice parameters calculated with LDA are $a = 3.87$ Å, $c = 3.42$ Å for $PbCuO_2$ with the $P4/mmm$ space group. For the $P4mm$ state, we have $a = 3.82$ Å, $c = 3.73$ Å. Pb, Cu, O1, and O2 atoms site at the $1b$ (1/2, 1/2, 0.5658), $1a$ (0, 0, 0.9972), and $2c$ (0, 1/2, 0.9434) (1/2, 0, 0.9434) Wyckoff positions respectively. The Cu-O-Cu bonding angle is about 168.02° due to the buckling of the CuO$_2$ plane in *P4mm* PbCuO$_2$. The bond length of Cu-O is 1.923 Å. The buckling parameter *d* is 0.20 Å. The calculated energy barrier is about 0.03 eV, which is much smaller than that of SnCuO$_2$.

### B. Electronic structures

In CaCuO$_2$, the stoichiometric Cu$^{2+}$ oxidation state leads a Cu $d^9$ configuration [2]. The paramagnetic (PM) electronic band structures calculated with HSE06 are given in Fig. 3(a). The most remarkable feature is that the Cu $3d_{x^2-y^2}$ band crosses the Fermi level. This half-filled band is the key to understand high-temperature superconductivity in cuprates [8]. The O 2*p* bands distribute from −7.5 to −2.5 eV. In Fig. 3(b) we show the PM electronic band structures of ferroelectric SnCuO$_2$. One can find that the Cu $3d_{x^2-y^2}$ band still crosses the Fermi level that is the same as CaCuO$_2$. However, the bandwidth is reduced by about 1 eV compared to the case in CaCuO$_2$, which indicates the influence of Sn on the *p-d* hybridization. Like CaCuO$_2$, SnCuO$_2$ also has a d$^9$ configuration. Along the Γ-Z direction, the Cu $3d_{x^2-y^2}$ band has weak dispersions showing a two-dimensional feature. Unlike CaCuO$_2$, there is also another band crossing the Fermi level enclosing holes around M, Z and R points in the Brillouin zone. At the M point, this band is contributed dominantly from Sn 5*s* and $p_z$ states with a little contribution from O 2$p_{x/y}$. At the Z point, the band is contributed from a mixture of Sn 5*s* and O 2$p_z$ states. From the formal charge of Sn$^{2+}$ one may expect that the inner 5*s* sate of Sn does not hybridize with other states. The Sn 5*s* sate mainly distributes from −10 to −7.5 eV as shown in Fig. 3(b). However, the appearance of Sn 5*s* sate around the Fermi level indicates its active behavior in SnCuO$_2$. The hybridization between the inner "core" *s* sate and O 2*p* states is known as the mechanism of the lone pair which is well studied in ferroelectrics [1]. As a result, there are highly asymmetric charge distributions around Sn as can be found in Fig. 1(d). The lone pair leads the buckling of Cu-O planes and thus induces ferroelectricity in SnCuO$_2$. Sn not only changes the atomic structures but also modifies the electronic structures around the Fermi level. Therefore, it is a nontrivial replacement of Ca by Sn.

To understand the bonding characters in SnCuO$_2$ more accurately, we obtain a tight-binding Hamiltonian based on maximally localized Wannier functions from the Wannier fitting of 12 bands around the Fermi energy corresponding to mainly Sn $s$ (1 state), Cu $d$ (5 states), and O $p$ (2 × 3 states) in the HSE06 calculations. The on-site energy difference between Cu $3d_{x^2-y^2}$ and O $2p_{x/y}$ orbitals is $\varepsilon_{dp} = \varepsilon_d - \varepsilon_p$. The Wannier fitting shows that $\varepsilon_{dp}$ is 2.0 eV. Previous works found that $\varepsilon_{dp}$ in CaCuO$_2$ is 2.7 eV. This energy is also known as the charge transfer energy in cuprates [2]. The on-site energy difference between Sn 5$s$ and O $2p_z$ orbitals is $\varepsilon_{sp} = \varepsilon_s - \varepsilon_p = 1.9$ eV. On the other hand, the amplitude of the hop between Cu $3d_{x^2-y^2}$ and O $2p_{x/y}$ orbitals is $t_{dp} = 1.5$ eV. The hop among O $2p_{x/y}$ orbitals is $t_{pp} = 0.6$ eV. We also find that the hop between Sn 5$s$ and O $2p_z$ orbitals is $t_{sp} = 0.3$ eV, which indicates a strong hybridization. This shows the active physical properties of the Sn 5$s$ orbital that is consistent with energy bands around the Fermi level discussed above. These tight-binding parameters give good descriptions of electronic structures for SnCuO$_2$ around the Fermi level.

In Fig. 4(a) we show the PM electronic band structures of polar phase PbCuO$_2$ calculated with HSE06. Different form SnCuO$_2$, there is only one band (Cu $3d_{x^2-y^2}$) crosses the Fermi level. Hence, PbCuO$_2$ still has a d$^9$ configuration. At the M and Z points, Pb 6$s$ state is just below the Fermi level. At the Γ point, Pb 6$s$ state locates at −7.5 eV, which is lower than that in SnCuO$_2$. We obtain the tight-binding Hamiltonian with Wannier fitting of 12 bands around the Fermi energy corresponding to mainly Pb $s$ (1 state), Cu $d$ (5 states), and O $p$ (2 × 3 states) in the HSE06 calculations. The on-site energy difference $\varepsilon_{dp}$ is 1.6 eV. The on-site energy difference between Pb 6$s$ and O $2p_z$ orbitals is $\varepsilon_{sp} = \varepsilon_s - \varepsilon_p = 0.4$ eV. The amplitude of $t_{dp}$ is 1.51 eV that is slightly smaller than that in SnCuO$_2$. The hop among O $2p_{x/y}$ orbitals is $t_{pp} =$ 0.5 eV. On the other hand, the hop between Pb 6*s* and O $2p_z$ orbitals is $t_{sp} = 0.9$ eV, which indicates a stronger hybridization than the hybridization between Sn 5*s* and O $2p_z$ orbitals in SnCuO$_2$. The Fermi surfaces for SnCuO$_2$ and PbCuO$_2$ calculated with the tight-binding Hamiltonian are shown in Fig. 4(b) and Fig. 4(c) respectively. Like CaCuO$_2$, the Fermi surface of PbCuO$_2$ is formed by the M-centered hole barrel. For SnCuO$_2$, Sn 5*s* sate around the Z point contributes the Fermi surface remarkably. In addition, the M-centered sphere is mainly Sn 5*s* in character.

### C. Magnetic order

Now we investigate the magnetic properties of the ferroelectric polar state SnCuO$_2$ with the *P4mm* symmetry calculated with the HSE06 functional, which is suitable for calculating electronic structures of strongly correlated copper oxides. The energy difference per unit cell between the PM and the ferromagnetic (FM) state is 141 meV. The energy difference per unit cell between the FM and C-AFM (C-type antiferromagnetic, AFM order within the CuO$_2$ plane and FM order between CuO$_2$ planes along the *c* direction) state is 307 meV. On the other hand, our HSE06 calculation shows that the G-AFM (G-type antiferromagnetic, AFM order within the CuO$_2$ plane and AFM order between CuO$_2$ planes along the *c* direction) state is the ground state of ferroelectric SnCuO$_2$. The energy of G-AFM state is 0.6 meV per unit cell lower than that of the C-AFM state. This small value indicates a weak interlayer magnetic coupling. According to Ref. [20], the intralayer nearest-neighbor exchange interaction constant $J_1$ and interlayer nearest-neighbor exchange interaction constant $J_{1z}$ could be obtained from:

$$J_1 = [E_{FM} - E_{C-AFM}]/4, \quad J_{1z} = [E_{C-AFM} - E_{G-AFM}]/2, \qquad (1)$$

where $E_{FM}$, $E_{C-AFM}$, and $E_{G-AFM}$ are the energies of FM, C-AFM, and G-AFM states, respectively. Then our theoretical results based on HSE06 are $J_1 = 76$ meV, $J_{1z} = 0.3$ meV, and the ratio $\frac{J_{1z}}{J_1} = 0.4\%$. For CaCuO$_2$, previous theoretical results with the B3LYP hybrid functional are $J_1 = 71$ meV, $J_{1z} = 0.8$ meV, and the ratio $\frac{J_{1z}}{J_1} = 1.1\%$ [20]. Previous experiments in Ref. [21] give $J_1 = 86$ meV, $J_{1z} = $ 0.4 meV. More recent experiments in Ref. [22] give $J_1 = 120$ meV. Compared to CaCuO$_2$, ferroelectric SnCuO$_2$ has a weaker interlayer magnetic coupling along the *c* direction. Theoretical magnetic moment on Cu in ferroelectric SnCuO$_2$ calculated with HSE06 is 0.64 $\mu_B$. The experimental magnetic moment on Cu in CaCuO$_2$ is 0.51 $\mu_B$.

From above discussions we know that the G-AFM state is the ground state of ferroelectric SnCuO$_2$. We study the electronic band structures by setting the G-type AFM spin configurations on Cu using a $\sqrt{2} \times \sqrt{2} \times 2$ supercell. The HSE06 electronic band structures are plotted in Fig. 5(a) with high-symmetry *k* points in the folded Brillouin zone. The Cu $3d_{x^2-y^2}$ bands split into lower and upper Hubbard bands separated by about 8 eV due to the strong correlation in polar SnCuO$_2$. The upper Hubbard Cu $3d_{x^2-y^2}$ band is just above the Fermi level and forms the conduction band. While the lower Hubbard Cu $3d_{x^2-y^2}$ band is located lower than O $2p$ bands. Sn $5s$ bands just below the Fermi level form the valence band. The energy gap is 0.37 eV, which is determined by the upper Hubbard Cu $3d_{x^2-y^2}$ band and top Sn $5s$ bands. Hence HSE06 results indicate that the ferroelectric SnCuO$_2$ is a charge transfer insulator with a narrow gap. As a ferroelectric material, the calculated electric polarization of polar sate SnCuO$_2$ is 0.57 C/m² that is comparable with typical ferroelectrics.

As discussed in Ref. [23], the band gap of transition-metal compounds (such as CuO and NiCl$_2$) is proportional to the charge transfer energy $\Delta$ ($d_i^n \to d_i^{n+1}L$ with *L* denotes a hole in the anion valence band) when the Coulomb interaction $U > \Delta$, which is a charge-transfer nature. Previous studies indicate that the energy gap of CaCuO$_2$ is 1.5 eV with the upper Hubbard Cu $3d_{x^2-y^2}$ band as the conduction band and the O $2p$ bands as the valence band. CaCuO$_2$ is a charge transfer type insulator [20]. For SnCuO$_2$ we also have $U > \Delta$ as can be found in Fig. 5(a). Due to the appearance of Sn $5s$ bands near the Fermi level, the energy gap of ferroelectric $SnCuO_2$ is much smaller than that of $CaCuO_2$. Different from these known charge transfer type insulators [23], the band gap of $SnCuO_2$ is not proportional to $\Delta$ but proportional to the gap $\Phi$ between the Cu $3d_{x^2-y^2}$ band and Sn $5s$ band around the Fermi level. Compounds similar to $SnCuO_2$ may be metals even though $U > \Delta > 0$, which is possible if $\Phi < 0$. Therefore, $SnCuO_2$ represents a type of charge transfer insulator that different from $CaCuO_2$. Schematic illustration of the charge transfer insulator $SnCuO_2$ and $PbCuO_2$ is shown in Fig. 5(b). From above discussions we know that Sn plays an important role in the electronic structures of $SnCuO_2$.

### D. Prospects

In experiments, $CaCuO_2$ can be grown on substrates such as $NdGaO_3$ and $SrTiO_3$ with the pulsed laser deposition (PLD) technique [24]. $SnCuO_2$ and $PbCuO_2$ predicted here may be synthesized with similar methods. Superconductivity appears in the hole and electron doped parent compounds such as $La_2CuO_4$ and $CaCuO_2$. According to the phase diagram of high-temperature superconductors, rich physical states emerge. Thus, in future experimental works, it is very interesting to study the doping properties of charge transfer multiferroics with a $d^9$ configuration designed in this work based on high-temperature superconductors. As multiferroics, it might be possible to control properties of doped phases by an applied electric field through magnetoelectric coupling effects. Multiferroics $SnCuO_2$ and $PbCuO_2$ designed in this work have the simplest crystal structures among materials containing copper-oxygen layers. We may make efforts to design other multiferroics with various crystal structures based on high-temperature superconductors with the lone pair mechanisms or other known and unknown physical mechanisms in future theoretical and experimental works.

### IV. CONCLUSIONS

In conclusion, we have designed multiferroics $SnCuO_2$ and $PbCuO_2$ with a $d^9$ configuration based on crystal structures of copper oxide $CaCuO_2$ with the help the lone pair mechanism. These multiferroics have the polar space group *P*4*mm* with the *P*4/*mmm* $CaCuO_2$ type crystal structure as their nonpolar high-symmetry paraelectric states. The active lone pair *s* electrons in Sn and Pb lead the buckling in the copper-oxygen layers and thus induce ferroelectricity. The electronic structures show that these multiferroics are charge transfer insulators with small band gaps. Multiferroics designed with the method proposed in this work retain some strongly correlated physical properties of the parent compound of copper oxide superconductors.

## ACKNOWLEDGMENTS

This work was supported by the Advanced Talents Incubation Program of the Hebei University (Grants No. 521000981423, No. 521000981394, No. 521000981395, and No. 521000981390), the Natural Science Foundation of Hebei Province of China (Grants No. A2021201001 and No. A2021201008), the National Natural Science Foundation of China (Grants No. 12104124 and No. 12274111), Scientific Research and Innovation Team of Hebei University (No. IT2023B03), and the high-performance computing center of Hebei University.

TABLE I. The lattice constants and atomic positions of nonpolar ($P4/mmm$) and polar ($P4mm$) states of $SnCuO_2$ and $PbCuO_2$ calculated with LDA.

| Crystals | Structural parameters | $P4/mmm$ | $P4mm$ |
|---|---|---|---|
| $SnCuO_2$ | $a$ | 3.88 Å | 3.78 Å |
| | $c$ | 3.21 Å | 3.98 Å |
| | Sn | $1d$ (1/2,1/2,1/2) | $1b$ (1/2,1/2,0.7016) |
| | Cu | $1a$ (0,0,0) | $1a$ (0,0,0.0771) |
| | O1 | $2f$ (0,1/2,0) | $2c$ (0,1/2,0.9926) |
| | O2 | (1/2,0,0) | (1/2,0,0.9926) |
| $PbCuO_2$ | $a$ | 3.87 Å | 3.82 Å |
| | $c$ | 3.42 Å | 3.73 Å |
| | Pb | $1d$ (1/2,1/2,1/2) | $1b$ (1/2,1/2,0.5658) |
| | Cu | $1a$ (0,0,0) | $1a$ (0,0,0.9972) |
| | O1 | $2f$ (0,1/2,0) | $2c$ (0,1/2,0.9434) |
| | O2 | (1/2,0,0) | (1/2,0,0.9434) |

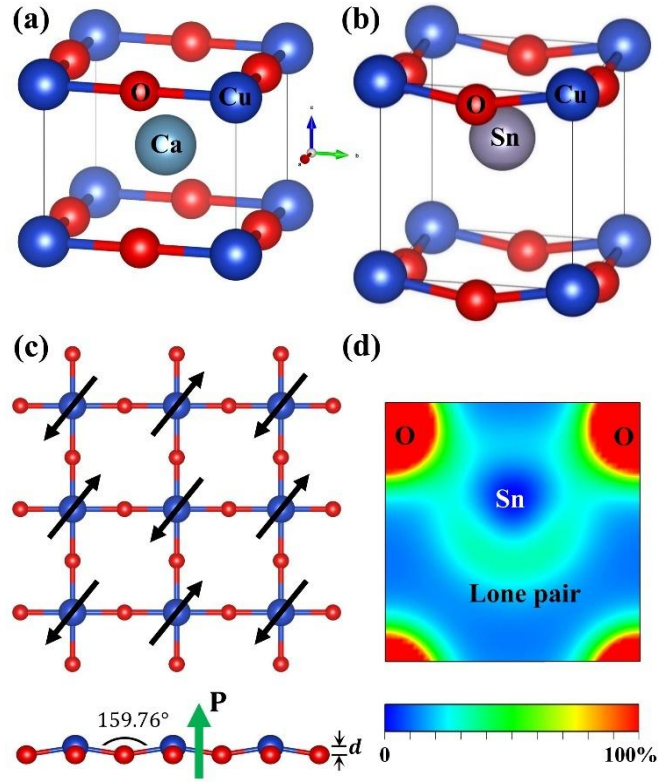

FIG. 1. The crystal structures of (a) CaCuO$_2$ with the $P$4/mmm symmetry (No. 123) and (b) the ferroelectric state SnCuO$_2$ with the $P$4mm symmetry (No. 99). (c) The top and side view of the buckled copper-oxygen layer in the ferroelectric state SnCuO$_2$. The black and green arrows indicate the spin and the electronic polarization respectively. The buckling parameter $d$ is defined as the distance between Cu and O in the buckled plane along the $c$ direction. (d) Electron densities of the (020) plane in ferroelectric state SnCuO$_2$.

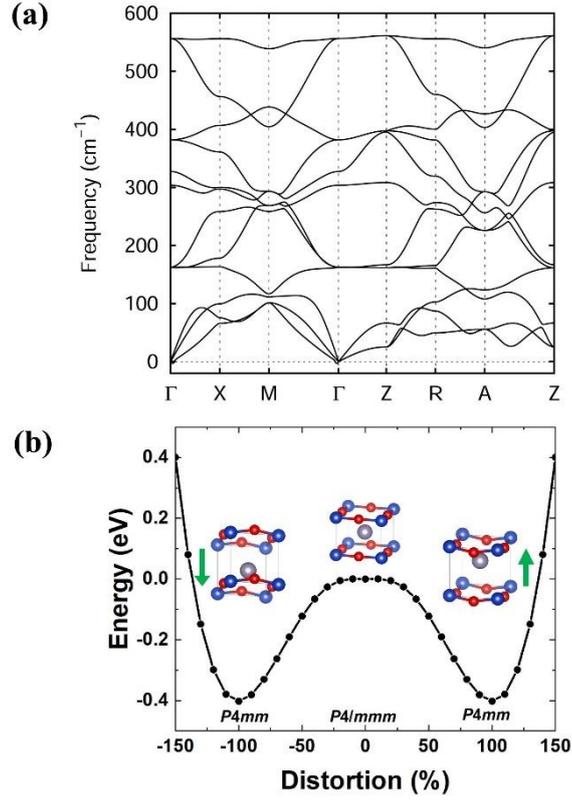

FIG. 2. (a) Phonon spectrum of ferroelectric state SnCuO$_2$ along the Γ(0, 0, 0)-X(0, 0.5, 0)-M(0.5, 0.5, 0)-Γ-Z(0, 0, 0.5)-R(0, 0.5, 0.5)-A(0.5, 0.5, 0.5)-Z directions. (b) The double-well energy of ferroelectric state SnCuO$_2$ as a function of polar distortion obtained by linear interpolation between the polar *P*4mm and nonpolar *P*4/mmm structures of SnCuO$_2$.

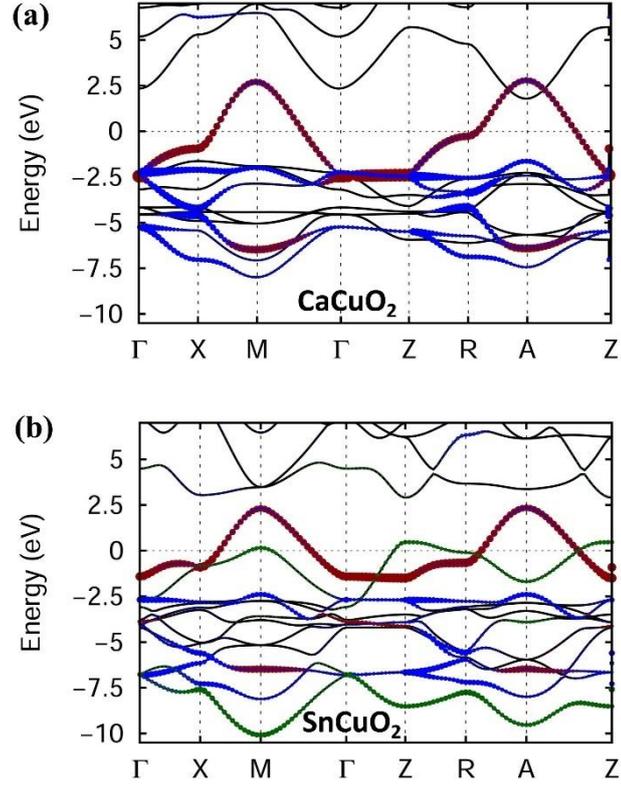

FIG. 3. HSE06 band structures of the nonmagnetic state in (a) CaCuO$_2$ and (b) the ferroelectric state SnCuO$_2$ along the Γ(0, 0, 0)-X(0, 0.5, 0)-M(0.5, 0.5, 0)-Γ-Z(0, 0, 0.5)-R(0, 0.5, 0.5)-A(0.5, 0.5, 0.5)-Z directions. The projected band structures of $d$ orbitals (dark red for $d_{x^2-y^2}$) in Cu, $2p$ orbitals (blue for $p_{x/y}$) in O, and $s$ orbitals (dark green) in Sn are shown.

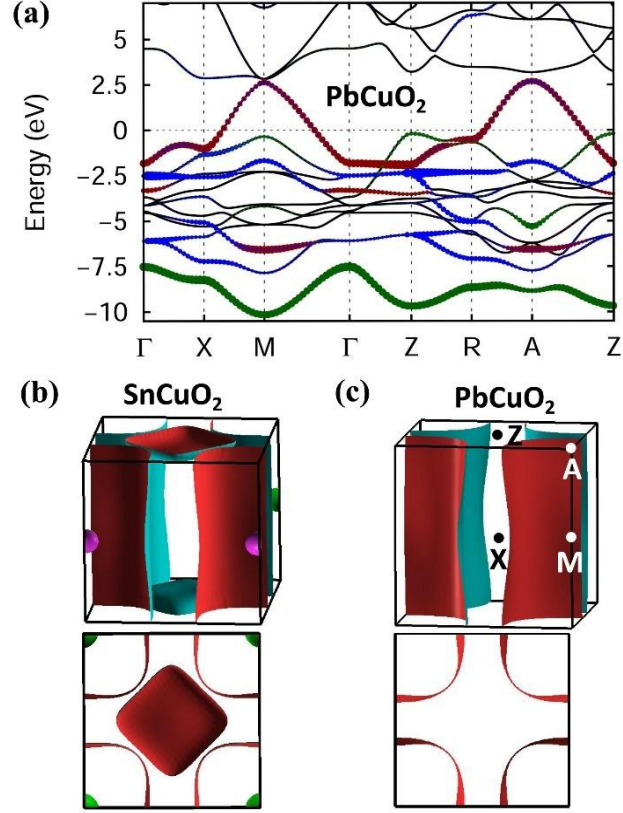

FIG. 4. (a) HSE06 band structures of the nonmagnetic state in ferroelectric state PbCuO$_2$. The color in band structures has the same meanings as in Fig. 3. Fermi surfaces of ferroelectric states (b) SnCuO$_2$ and (c) PbCuO$_2$ are shown.

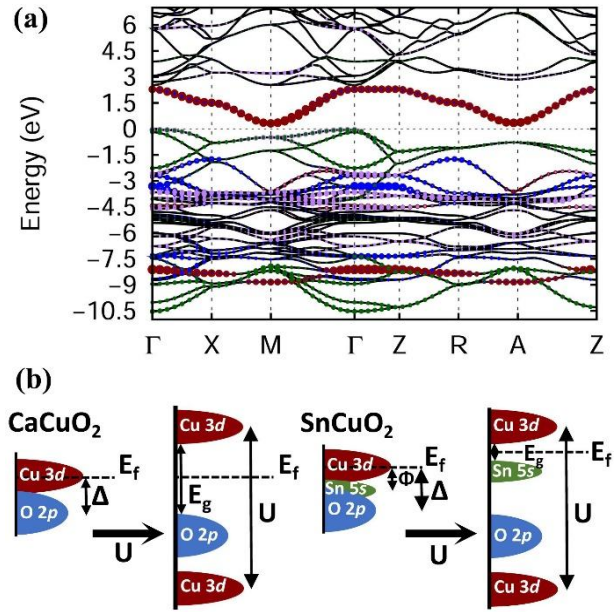

FIG. 5. (a) HSE06 band structures of G-type of antiferromagnetic state for a $\sqrt{2} \times \sqrt{2} \times 2$ supercell in the ferroelectric state SnCuO$_2$. The notation Γ, X, M, Z, R, and A is in the folded magnetic Brillouin zone. (b) Schematic illustration of the charge transfer insulator CaCuO$_2$ and SnCuO$_2$ generated by the $d$-site interaction effect. $E_g$ and $E_f$ denote the charge gap and Fermi level respectively.